# Exploiting network optimization stability for enhanced PET image denoising using deep image prior

**Fumio Hashimoto[1,2,*,**], Kibo Ote[1], Yuya Onishi[1], Hideaki Tashima[2], Go Akamatsu[2], Yuma Iwao[2], Miwako Takahashi[2], and Taiga Yamaya[2, **]**

[1] Central Research Laboratory, Hamamatsu Photonics K. K., 5000 Hirakuchi, Hamana-ku, Hamamatsu 434-8601, Japan

[2] Institute for Quantum Medical Science, National Institutes for Quantum Science and Technology (QST), 4-9-1 Anagawa, Inage-ku, Chiba 263-8555, Japan

[*] Current affiliation: J. Crayton Pruitt Family Department of Biomedical Engineering, University of Florida, Gainesville, 32611, FL, USA

[**] Author to whom any correspondence should be addressed.

E-mail: fumio.hashimo@ufl.edu; yamaya.taiga@qst.go.jp

## Abstract

[Objective]Positron emission tomography (PET) is affected by statistical noise due to constraints on tracer dose and scan duration, impacting both diagnostic performance and quantitative accuracy. While deep learning-based PET denoising methods have been used to improve image quality, they may introduce over-smoothing, which can obscure critical structural details and compromise quantitative accuracy. We propose a method for making a deep learning solution more reliable and apply it to the conditional deep image prior (DIP).

[Approach]We introduce the idea of *stability information* in the optimization process of conditional DIP, enabling the identification of unstable regions within the network's optimization trajectory. Our method incorporates a stability map, which is derived from multiple intermediate outputs of a moderate neural network at different optimization steps. The final denoised PET image is then obtained by computing a linear combination of the DIP output and the original reconstructed PET image, weighted by the stability map.

[Main results]We employed eight high-resolution brain PET datasets for comparison. Our method effectively reduces background noise while preserving small structure details in brain [18F]FDG PET images. Comparative analysis demonstrated that our approach outperformed existing methods in terms of peak-to-valley ratio and background noise suppression across various low-dose levels. Additionally, region-of-interest analysis confirmed that the proposed method maintains quantitative accuracy without introducing under- or over-estimation. Furthermore, we applied our method to full-dose PET data to assess its impact on image quality. The results revealed that the proposed method significantly reduced background noise while preserving the peak-to-valley ratio at a level comparable to that of unfiltered full-dose PET images.

[Significance]The proposed method introduces a robust approach to deep learning-based PET denoising, enhancing its reliability and preserving quantitative accuracy. Furthermore, this strategy has the potential to advance performance in high-sensitivity PET scanners, demonstrating that deep learning can extend PET imaging capabilities beyond low-dose applications.

## 1. Introduction

Positron emission tomography (PET) is a widely used medical imaging modality for assessing pharmacokinetics in a living body. PET images often suffer from statistical noise due to inherent limitations in tracer dose and scan duration, compromising both diagnostic accuracy and quantitative accuracy. Gaussian filtering is commonly employed for post-denoising in clinical applications, but while effective in reducing noise, it also tends to blur critical structural details. To address these limitations, advanced PET image denoising techniques, such as bilateral [1], nonlocal means [2], guided [3], and block-matching [4] filterings, have been developed to suppress statistical noise while preserving spatial resolution.

Recent studies have demonstrated the potential of deep learning in PET imaging [5, 6], demonstrating superior denoising performance and improved spatial resolution preservation compared to conventional filtering techniques. A typical deep learning-based PET denoising framework relies on supervised learning, which requires large-scale training datasets comprising high-quality or high-dose PET images and their corresponding low-quality or low-dose counterparts. However, the effectiveness of these models depends heavily on the quality and quantity of the training data. Acquiring such datasets presents significant challenges due to confidentiality concerns, ethical restrictions, and the inherent variability in PET data. Variations in PET scanners, tracers, and disease-related racial disparities further complicate dataset preparation, making it impractical to construct a fully representative dataset that captures the entire representation (or domain) of PET images.

An effective strategy to address these challenges is to employ an unsupervised learning approach that performs denoising without requiring high-quality PET image data. One such method, Noise2Noise [7], learns the denoising task using only low-quality data pairs and has been successfully applied to PET image denoising [8, 9]. Noise2Void [10, 11] and deep image prior (DIP) [12, 13] achieve denoising using only the target (noisy) image itself. DIP is particularly well suited for PET imaging applications and has significantly expanded its scope to include PET image denoising and reconstruction [14-17]. Early implementations of DIP for PET image denoising incorporated conditional priors such as X-ray computed tomography (CT), magnetic resonance (MR) images [18], and static PET images [19] into commonly used convolutional neural networks (CNNs), including the U-Net architecture. This approach, known as conditional DIP, enhances PET image denoising performance compared to the original DIP framework, which uses a random noise distribution rather than conditional priors. Further advancements in neural network architectures [20, 21] and the integration of pre-trained models [22, 23] independently contributed to improved PET image denoising performance within the DIP framework.

While these unsupervised PET image denoising methods effectively suppress statistical noise, the resulting denoised images often appear more blurred than those generated by state-of-the-art supervised learning approaches, such as generative adversarial networks [24, 25] and diffusion models [26]. Additionally, deep learning-based denoising may cause over-smoothing, leading to the loss of critical structural details and compromising quantitative accuracy. Therefore, a more sophisticated deep learning-based algorithm is required to preserve essential structural details while ensuring reliable and quantitatively accurate PET images.

In this work, we propose a novel method for making a deep learning solution more reliable and apply it to the conditional DIP framework for PET image denoising. Specifically, we introduce the idea of *stability information* within the optimization process of conditional DIP denoising to identify unstable regions during neural network optimization. By leveraging this stability information, we assess the denoising performance of deep learning-generated results. To validate our approach, we quantitatively compare the proposed denoising method with ordered subset expectation maximization (OSEM) and the original DIP framework in real brain PET imaging using $[^{18}F]$FDG.

## 2. Methodology

### 2.1 Conditional DIP

The DIP is an optimization method for image restoration tasks that exploit the structural or inductive biases of CNN structures, which inherently favor image generation. The original DIP, introduced by Ulyanov et al [12, 13], uses a single pair of degraded image as a target data and a random noise distribution as a CNN input. Instead, the conditional DIP, a modified version of the original DIP framework, incorporates a conditional prior $g$, such as X-ray CT, MR, or static PET images, to obtain a denoised image $x^*$, as follows [18, 19],

$$\theta^* = \arg\min_{\theta}\|x_0 - f(\theta|g)\|,$$
$$x^* = f(\theta^*|g), \tag{1}$$

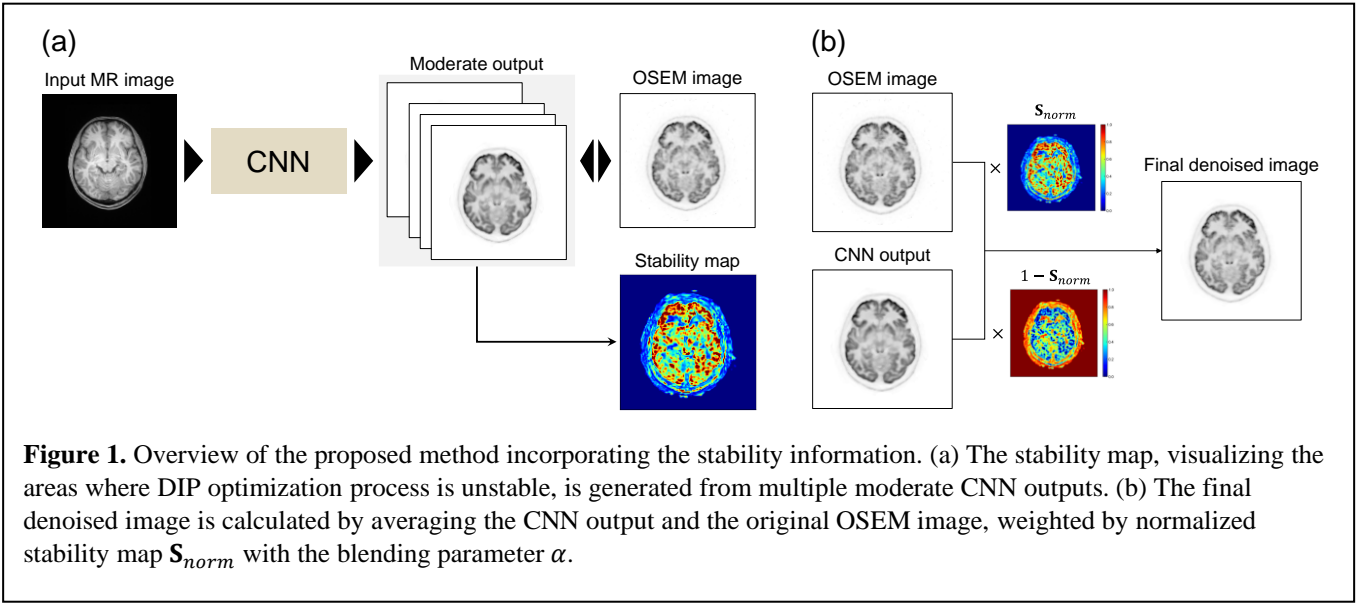

**Figure 1.** Overview of the proposed method incorporating the stability information. (a) The stability map, visualizing the areas where DIP optimization process is unstable, is generated from multiple moderate CNN outputs. (b) The final denoised image is calculated by averaging the CNN output and the original OSEM image, weighted by normalized stability map $\mathbf{S}_{norm}$ with the blending parameter $\alpha$.

where $f$ is the neural network with weights $\theta$, and $\boldsymbol{x}_0$ represents the noisy PET image. The conditional DIP offers an advantage by positioning an initial point of an optimization trajectory in a manifold closer to the ground truth compared to the original DIP. Consequently, the conditional DIP achieves more stable optimization and produces higher-quality PET images [18, 20].

### 2.2 Proposed method

In this study, we introduce a reliable PET image denoising approach based on the conditional DIP framework. An overview of the proposed method is illustrated in Figure 1.

First, conditional DIP denoising is performed using an MR image as the conditional prior and the corresponding noisy PET image—reconstructed using the OSEM algorithm—as the CNN input. Subsequently, a stability map $\mathbf{S}$ is computed from multiple moderate CNN outputs, sampled at fixed intervals $M$, to generate the standard deviation image as follows:

$$\boldsymbol{S} = Std\big[f\big(\theta_{n|n\%M=0,n\geq m}\big|\boldsymbol{g}\big)\big], \tag{2}$$

where $Std[\cdot]$ denotes the standard deviation operator, and $\theta_n$ represents trainable weights at epochs $n$. $m$ is a hyperparameter introduced to ensure that weight sampling begins only after the optimization process has progressed beyond its initial, rapidly changing phase. By restricting the calculation of $\mathbf{S}$ to epochs $n \geq m$, weight samples are obtained from a more stable region of the optimization trajectory, providing a more reliable assessment of the model's performance and variability. In other words, early optimization steps (i.e., those for which $n < m$) are omitted, as they typically exhibit large fluctuations in the parameter space. The fixed interval $M$ is used to maintain statistical independence between the sampled CNN outputs.

The stability map provides a visualization of regions where DIP optimization exhibits instability. Consequently, the final denoised PET image $\boldsymbol{x}$ is computed as a linear combination of the CNN-denoised output and the original OSEM image $\boldsymbol{x}_0$, weighted by a normalized stability map $\mathbf{S}_{norm}$ at voxel index $j$, as follows:

$$x = S_{norm}x_0 + (1 - S_{norm})f(\theta^*|g),$$
$$(S_{norm})_j = min\left(\alpha\frac{S_j}{\bar{S}}, 1\right), \tag{3}$$

where $\bar{S}$ is the mean value of the stability map $\mathbf{S}$, and $\alpha$ is a hyperparameter that adjusts the degree of blending. Note that the multiplication of the vectors is computed element-wise.

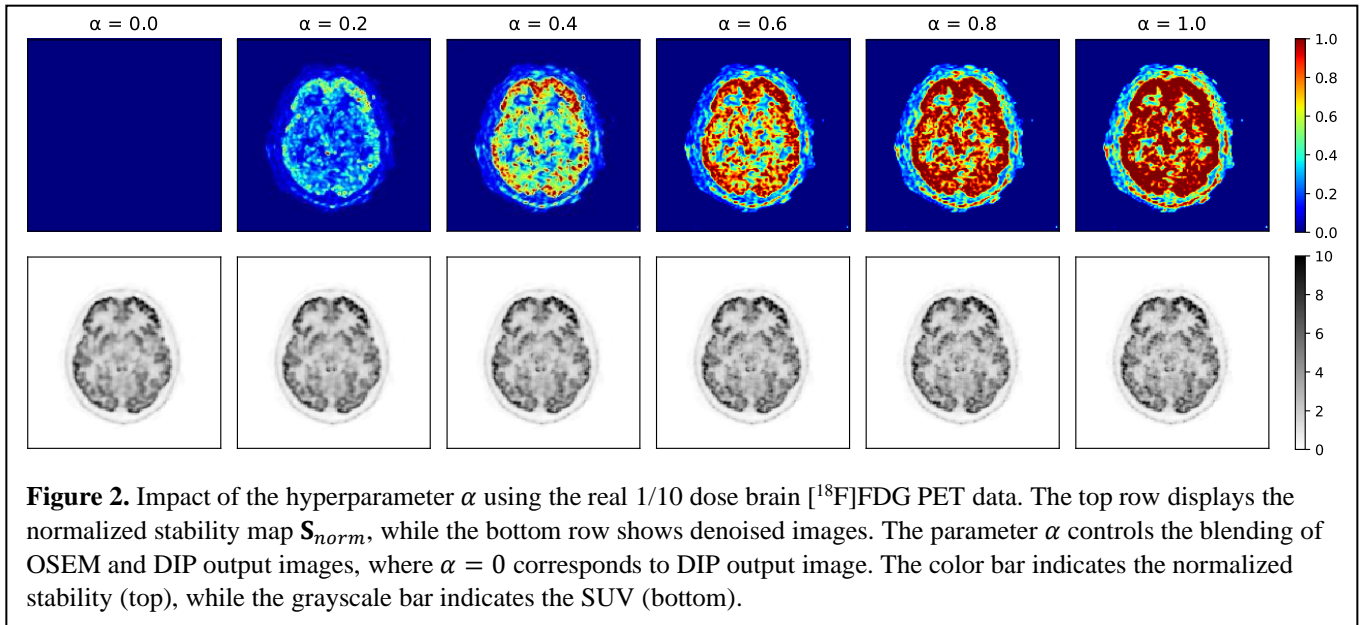

**Figure 2.** Impact of the hyperparameter $\alpha$ using the real 1/10 dose brain [$^{18}$F]FDG PET data. The top row displays the normalized stability map $\mathbf{S}_{norm}$, while the bottom row shows denoised images. The parameter $\alpha$ controls the blending of OSEM and DIP output images, where $\alpha = 0$ corresponds to DIP output image. The color bar indicates the normalized stability (top), while the grayscale bar indicates the SUV (bottom).

The proposed method is designed to increase the contribution of the OSEM image in regions (voxels) where stability values are low, corresponding to a high sample standard deviation in DIP outputs. This ensures that the algorithm relies on OSEM to mitigate potential artifacts from the DIP output. Conversely, voxels with high stability values predominantly adopt the DIP output, improving noise characteristics. This weighted averaging strategy enhances the overall reliability of the reconstructed PET image. As a result, the proposed method effectively balances the robustness of the OSEM algorithm with denoising capability and detail recovery of the conditional DIP denoising framework induced by MR image prior.

We employ a typical 3D U-Net architecture, identical to the one used in [23] in this study. The limited-memory Broyden–Fletcher–Goldfarb–Shanno algorithm [27], a quasi-Newton optimization method that utilizes second-order gradient information and converges faster than first-order optimizers, was selected as the optimizer. In the proposed method, the number of epochs, $M$ and $m$, were set to 200, 5, and 100, respectively. The experiments were conducted using PyTorch 1.12.1 and an NVIDIA A100 graphics card with 80 GB of memory.

## 3. Experimental setup

### 3.1 Dataset

We employed brain PET images from eight healthy male volunteers (aged 22–47 years), acquired using a brain-dedicated PET scanner (VRAIN$^{TM}$, ATOX Co., Ltd., Japan) [28, 29], along with separately acquired MR images. None of the participants had a history of brain injury, psychiatric disorders, or abnormal findings on MR images. A dose of [$^{18}$F]FDG ($289 \pm 25$ MBq) was administered following a minimum fasting period of six hours. Participants' forehead and chin were stabilized using fixation bands, and a 30-minute PET scan was conducted after 104 (range, 100–113) minutes post-administration. We simulated low-dose PET data for performance evaluation by randomly downsampling the acquired emission list-mode data to 1/10, 1/5, and 1/2 of the original events. Additionally, we evaluated the full-dose PET data to determine whether the proposed method further enhances PET image quality. The reconstructed image size was $136 \times 136 \times 112$ voxels with $2.0 \times 2.0 \times 2.0$ mm$^3$.

The study protocol adhered to the principles outlined in the Declaration of Helsinki and was approved by the Institutional Review Board of QST Hospital, Japan. Written informed consent was obtained from all participants prior to their inclusion in the study. Additionally, the study was registered with the University Hospital Medical Information Network (UMIN000051244).

### 3.2 Evaluation

We compared the performance of the proposed denoising method with that of OSEM with Gaussian post-filtering and conditional DIP [18]. For a fair comparison, the proposed and conditional DIP methods employed the same network architecture. The full width at half maximum (FWHM) of the Gaussian post-filtering was varied from 0.0 to 4.0 voxels in increments of 0.5 voxels. Conditional DIP denoising was evaluated every 10 epochs from 110 to 200, while the α values of the proposed method ranged from 0.0 to 1.0 in increments of 0.1.

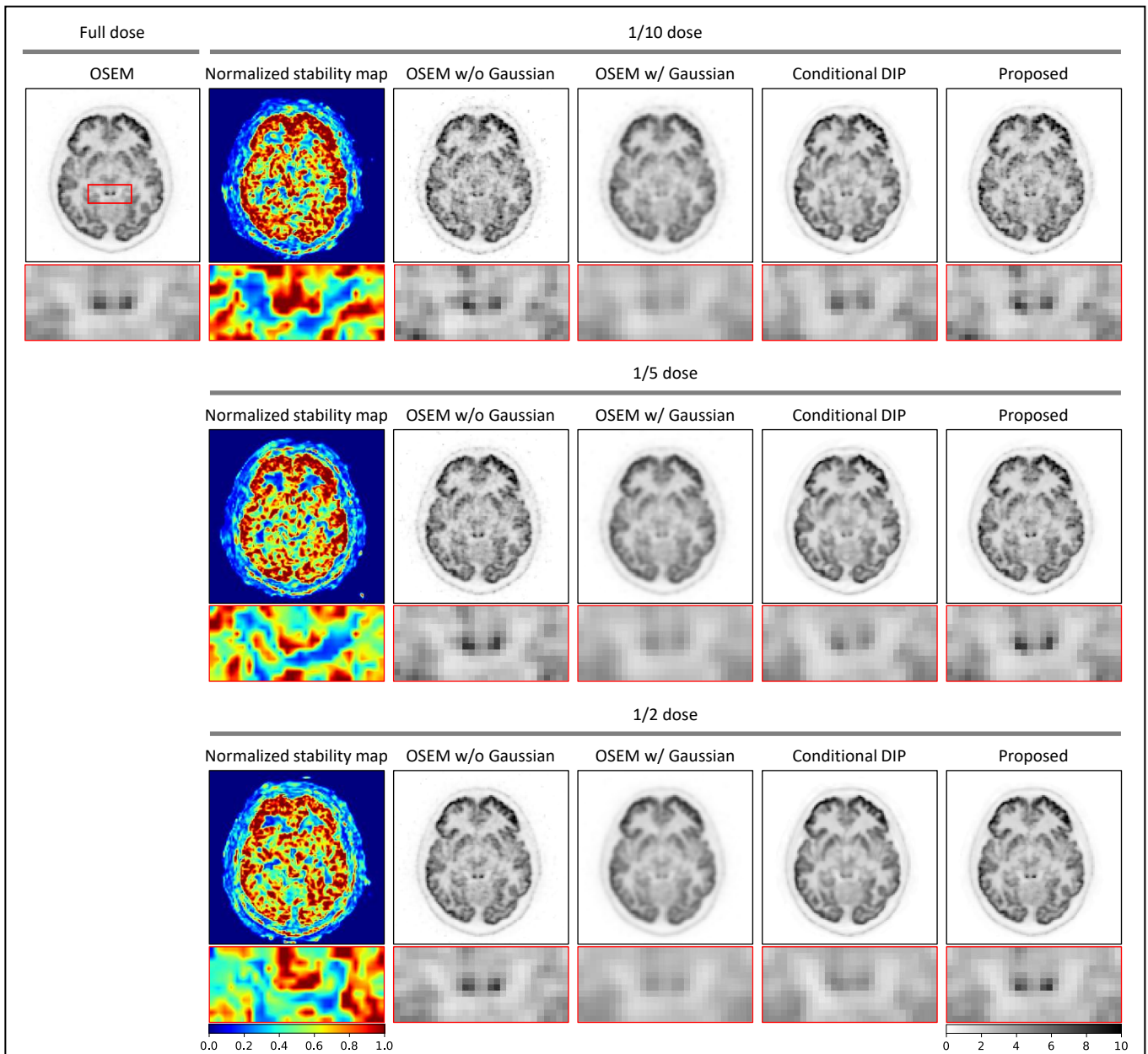

**Figure 3.** Denoising results for real brain [$^{18}$F]FDG PET data using different methods. The reference image (full-dose OSEM) is shown for comparison, followed by low-dose results from the normalized stability map, OSEM without Gaussian post-filtering, OSEM with Gaussian post-filtering (FWHM = 2.0 voxels), conditional DIP (200 epochs), and the proposed method ($\alpha = 0.6$) (from left to right). The rows correspond to different low-dose levels: 1/10, 1/5, and 1/2. The magnified images of the red squared box are shown in each corresponding bottom row. The color bar indicates the normalized stability (lower left), while the grayscale bar indicates the SUV (lower right).

To assess PET image denoising performance, we calculated the peak-to-valley ratio in the small nuclei and inferior colliculus in the midbrain and measured the standard deviation of the white matter (background). Additionally, we performed a region-of-interest (ROI) analysis using PMOD software (PMOD Technologies Ltd, Zurich, Switzerland) with the automated anatomical labeling ROI template. Standard uptake values (SUVs) were calculated for ROIs in the cerebellar cortex, frontal cortex, medial temporal cortex, parietal cortex, posterior cingulate and precuneus, striatum, and thalamus. For this analysis, we applied an FWHM of 1.5 voxels for Gaussian post-filtering, 200 epochs for conditional DIP denoising, and $\alpha$ value of 0.6 for the proposed method.

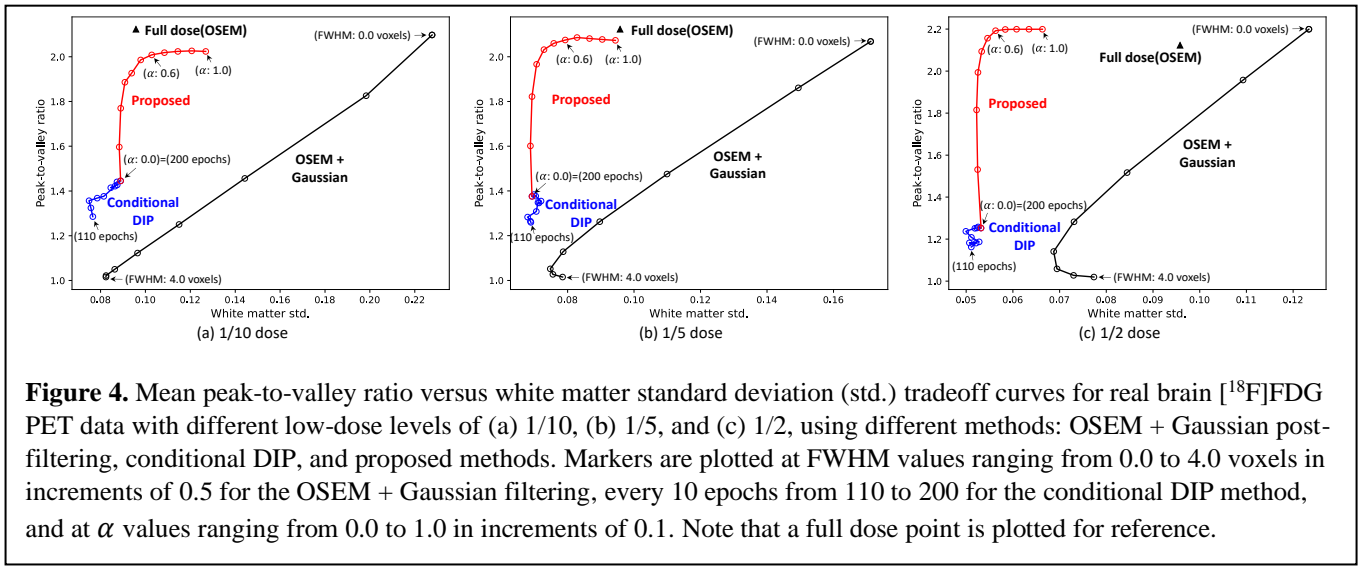

**Figure 4.** Mean peak-to-valley ratio versus white matter standard deviation (std.) tradeoff curves for real brain [$^{18}$F]FDG PET data with different low-dose levels of (a) 1/10, (b) 1/5, and (c) 1/2, using different methods: OSEM + Gaussian post-filtering, conditional DIP, and proposed methods. Markers are plotted at FWHM values ranging from 0.0 to 4.0 voxels in increments of 0.5 for the OSEM + Gaussian filtering, every 10 epochs from 110 to 200 for the conditional DIP method, and at $\alpha$ values ranging from 0.0 to 1.0 in increments of 0.1. Note that a full dose point is plotted for reference.

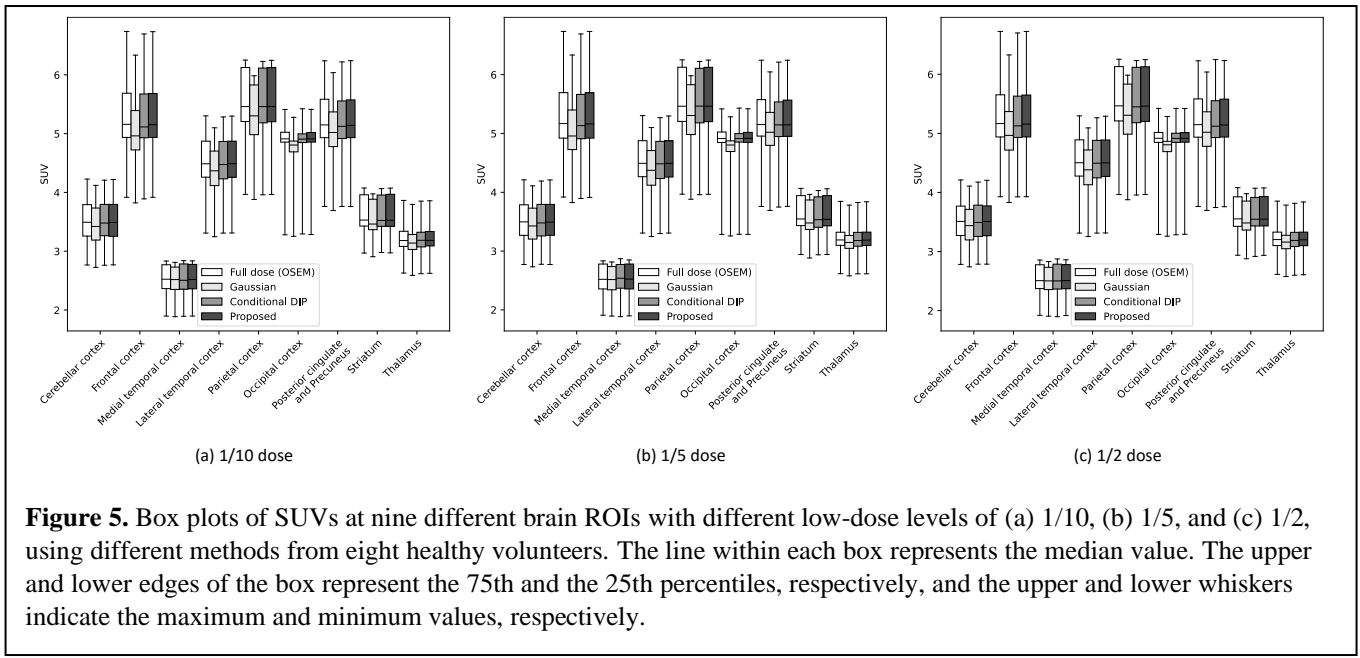

**Figure 5.** Box plots of SUVs at nine different brain ROIs with different low-dose levels of (a) 1/10, (b) 1/5, and (c) 1/2, using different methods from eight healthy volunteers. The line within each box represents the median value. The upper and lower edges of the box represent the 75th and the 25th percentiles, respectively, and the upper and lower whiskers indicate the maximum and minimum values, respectively.

## 4. Results

Figure 2 illustrates the effect of the hyperparameter $\alpha$ on real 1/10-dose brain [$^{18}$F]FDG PET data. We observed that $\alpha$, which controls the blending of OSEM and DIP output images, functions as a smoothing parameter. The stability map revealed high stability in homogeneous regions such as white matter and low stability in complex or small structures, including the cortex.

Figure 3 presents the denoising results for real brain [$^{18}$F]FDG PET data at dose levels of 1/10, 1/5, and 1/2 using different methods: OSEM with Gaussian post-filtering, conditional DIP, and the proposed method. The results demonstrate that the proposed denoising method effectively reduces statistical noise in white matter regions while preserving hot spot structures, such as the small nuclei and inferior colliculus in the midbrain. In contrast, the conditional DIP method tends to produce excessively smoothed and low-contrast images, with this tendency becoming more pronounced as the dose increases.

Figure 4 shows the mean peak-to-valley ratio and white matter standard deviation tradeoff curves for real brain [$^{18}$F]FDG PET data at dose levels of 1/10, 1/5, and 1/2, derived from eight healthy male volunteers. The proposed method consistently achieved both a higher peak-to-valley ratio and a lower noise level than the other methods across all dose levels. Figure 5 presents box plots of SUVs for nine different brain ROIs obtained using different methods. Compared to the full-dose reference, the proposed method maintains similar SUV distribution patterns. In contrast, the OSEM with Gaussian post-filtering and conditional DIP methods exhibit decreased SUVs in most brain regions.

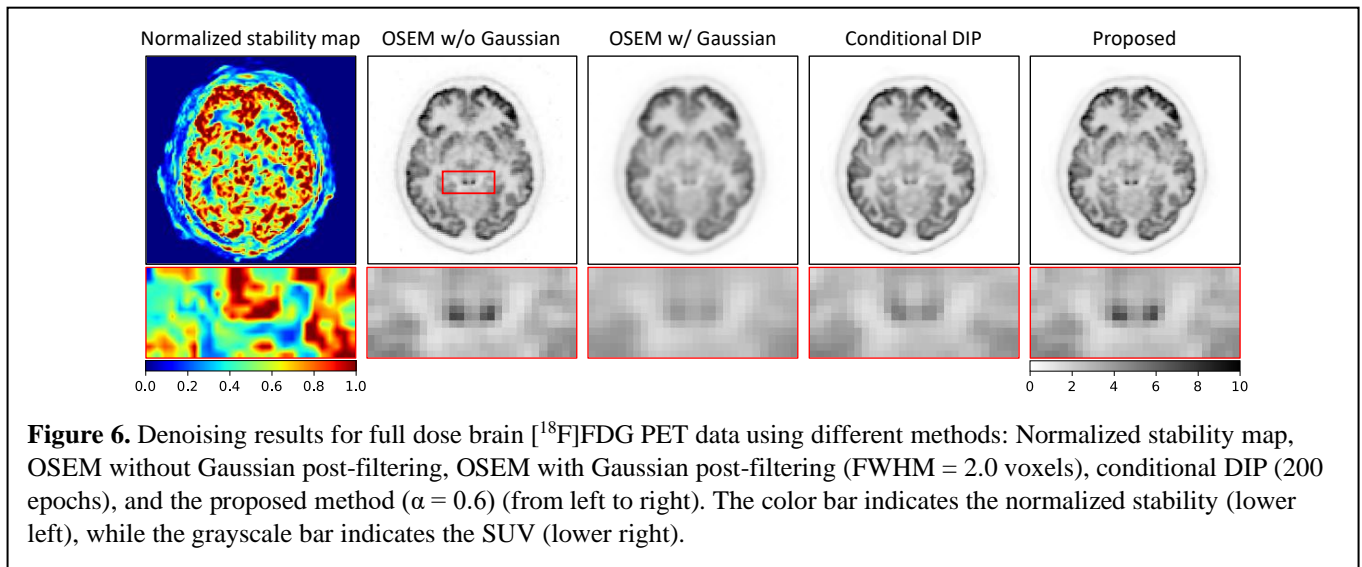

**Figure 6.** Denoising results for full dose brain [$^{18}$F]FDG PET data using different methods: Normalized stability map, OSEM without Gaussian post-filtering, OSEM with Gaussian post-filtering (FWHM = 2.0 voxels), conditional DIP (200 epochs), and the proposed method ($\alpha = 0.6$) (from left to right). The color bar indicates the normalized stability (lower left), while the grayscale bar indicates the SUV (lower right).

Figure 6 presents the denoising results for real brain [$^{18}$F]FDG PET data at the full-dose level, comparing OSEM with Gaussian post-filtering, conditional DIP, and the proposed method. The proposed method exhibits excellent denoising performance, even on high-quality full-dose PET images, whereas the conditional DIP method tends to produce over-smoothed PET images with reduced texture detail. The quantitative evaluation results further support these observations. Figure 7 illustrates the mean peak-to-valley ratio and white matter standard deviation tradeoff curves, demonstrating the superior performance of the proposed method. Figure 8 presents box plots of SUVs across nine different brain ROIs, further validating the effectiveness of the proposed method in preserving quantitative accuracy.

## 5. Discussion

In this study, we introduced a novel method for making a deep learning solution more reliable and apply it to the conditional DIP framework for PET image denoising. To identify unstable regions during network optimization, we computed a standard deviation map from multiple moderate CNN outputs and utilized it as stability information.

The denoising results for low-dose PET data demonstrated that the proposed method preserved fine structural details more effectively than the conditional DIP method, which tends to produce over-smoothed, low-texture PET images. Notably, the proposed method successfully restored the peak-to-valley ratio to full-dose levels while simultaneously reducing background statistical noise. This improvement occurs because the proposed method selectively incorporates OSEM image information in regions where the stability map values are low—areas where conditional DIP denoising would otherwise blur fine structures.

DIP denoising separates signals in noisy images by leveraging differences in the convergence rates of signals and noise, an effect arising from the structural or inductive bias of the CNN architecture. However, this approach struggles to recognize small structures spanning only a few voxels, often misidentifying them as noise. The proposed method addresses this limitation by employing a stability map to visualize regions where DIP denoising fails to differentiate small structures from noise. By integrating this stability information into the denoising process, the proposed method effectively mitigates signal loss, ensuring better preservation of fine structural details.

We conducted a full-dose PET evaluation to determine whether the proposed method further enhances PET image quality. The results demonstrated that the proposed method successfully restored the peak-to-valley ratio at a level comparable to the unfiltered full-dose PET images using the OSEM algorithm while effectively reducing background noise. Additionally, ROI analysis confirmed that the proposed method did not introduce under- or over-estimation. These findings suggest that the proposed approach has the potential to advance imaging performance in state-of-the-art high-sensitivity PET scanners, enabling deep learning-based denoising to extend PET imaging capabilities beyond low-dose applications.

Although the proposed method utilizes conditional DIP denoising as its framework, diffusion models are increasingly being adopted for PET image denoising due to their superior performance [26, 30,31]. The algorithmic structure of diffusion models allows for the straightforward calculation of standard deviation maps through multiple sampling. Given this compatibility, the proposed procedure can be seamlessly integrated into diffusion models for enhanced denoising performance.

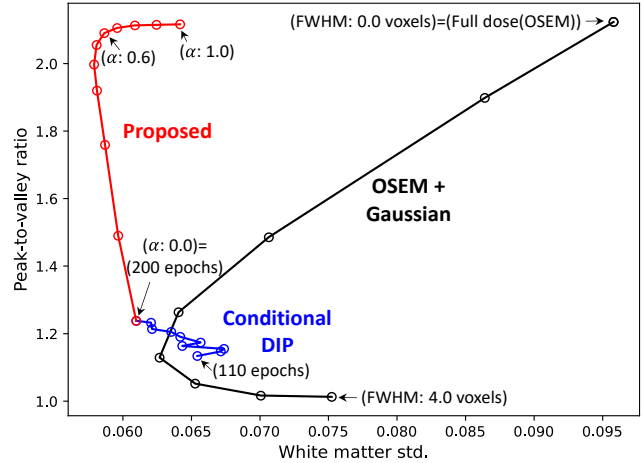

**Figure 7.** Mean peak-to-valley ratio versus white matter standard deviation (std.) tradeoff curves for full dose brain [$^{18}$F]FDG PET data using different methods: OSEM + Gaussian filtering, conditional DIP, and proposed methods. Markers are plotted at FWHM values ranging from 0.0 to 4.0 voxels in increments of 0.5 for the OSEM + Gaussian filtering, every 10 epochs from 110 to 200 for the conditional DIP method, and at $\alpha$ values ranging from 0.0 to 1.0 in increments of 0.1.

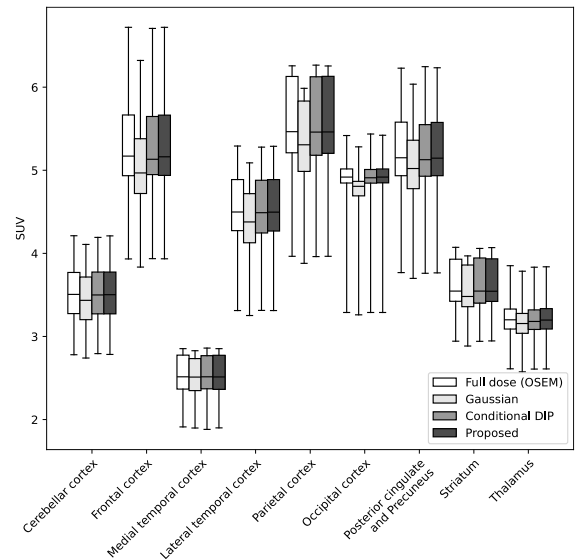

**Figure 8.** Box plots of SUVs at nine different brain ROIs for full dose [$^{18}$F]FDG PET data using different methods from eight healthy volunteers. The line within each box represents the median value. The upper and lower edges of the box represent the 75th and the 25th percentiles, respectively, and the upper and lower whiskers indicate the maximum and minimum values, respectively.

One challenge associated with conditional DIP is the potential bias introduced by mismatches between prior images and PET images—for example, when tumors appear in PET but are absent in MR images [23, 32]. In our experiments, we observed that such mismatches contributed to slower convergence and higher sample standard deviations in the stability map, particularly in regions such as the small nuclei in the midbrain (Figure 2). The proposed method is expected to mitigate this mismatch-induced bias due to its ability to adaptively incorporate unbiased reconstructed PET images.

A primary limitation of this study is that it was restricted to brain [$^{18}$F]FDG PET data from healthy volunteers. However, previous research has demonstrated that the conditional DIP method is applicable to a wide range of PET datasets for image denoising [18,20–23]. As part of future work, we plan to broaden our evaluation by incorporating diverse datasets, including different PET tracers, organs, scanners, and disease conditions.

## 6. Conclusion

In this study, we introduced a novel method for making a deep learning solution more reliable and apply it to the conditional DIP framework for PET image denoising. Experimental evaluations demonstrated that the proposed DIP-based denoising method outperformed Gaussian post-filtering and conditional DIP denoising, particularly in low-dose PET imaging scenarios. Furthermore, quantitative ROI analysis confirmed that the proposed method preserves quantitative accuracy without introducing biases. These findings suggest that the proposed stability information-based approach represents a promising advancement in PET image denoising. Additionally, the proposed method effectively reduced background noise while maintaining the peak-to-valley ratio at a level comparable to the unfiltered full-dose PET images using the OSEM algorithm. ROI analysis further confirmed that the method did not cause under- or over-estimation, ensuring its quantitative reliability. These findings indicate that the proposed procedure has the potential to enhance imaging performance in state-of-the-art high-sensitivity PET scanners, enabling deep learning-based techniques to extend PET imaging capabilities beyond low-dose applications.

## Acknowledgments

This work was supported by JSPS KAKENHI Grant Number JP22K07762.

## Data availability statement

Clinical data cannot be publicly shared due to sensitive personal information.

## References

1    Hofheinz F, Langner J, Petr J, Beuthien-Baumann B, Oehme L and Steinbach J 2011 Suitability of bilateral filtering for edge-preserving noise reduction in PET *EJNMMI Res.* **1** 23
2    Arabi H and Zaidi H 2021 Non-local mean denoising using multiple PET reconstructions *Ann. Nucl. Med.* **35** 176–86
3    Hashimoto F, Ohba H, Ote K and Tsukada H 2018 Denoising of dynamic sinogram by image-guided filtering for positron emission tomography *IEEE Trans. Radiat. Plasma Med. Sci.* **2** 541–8
4    Ote K, Hashimoto F, Kakimoto A, Isobe T, Inubushi T, Ota R, ... and Ouchi Y 2020 Kinetics-induced block matching and 5-D transform domain filtering for dynamic PET image denoising *IEEE Trans. Radiat. Plasma Med. Sci.* **4** 720–8
5    Bousse A, Kandarpa V S S, Shi K, Gong K, Lee J S, Liu C and Visvikis D 2024 A review on low-dose emission tomography post-reconstruction denoising with neural network approaches *IEEE Trans. Radiat. Plasma Med. Sci.* **8** 1–9
6    Hashimoto F, Onishi Y, Ote K, Tashima H, Reader A J and Yamaya T 2024 Deep learning-based PET image denoising and reconstruction: a review *Radiol. Phys. Technol.* **17** 24–46
7    Lehtinen J, Munkberg J, Hasselgren J, Laine S, Karras T, Aittala M and Aila T 2018 Noise2Noise: Learning Image Restoration without Clean Data *arXiv* 1803.04189
8    Yie S Y, Kang S K, Hwang D and Lee J S 2020 Self-supervised PET denoising *Nucl. Med. Mol. Imaging* **54** 299–304
9    Kang S K, Yie S Y and Lee J S 2021 Noise2Noise improved by trainable wavelet coefficients for PET denoising *Electronics* **10** 1529
10   Krull A, Buchholz T-O and Jug F 2019 Noise2Void—Learning denoising from single noisy images *Proc. IEEE/CVF Conf. Comput. Vis. Pattern Recognit.* pp 2129–37
11   Song T A, Yang F and Dutta J 2021 Noise2Void: unsupervised denoising of PET images *Phys. Med. Biol.* **66** 214002
12   Ulyanov D, Vedaldi A and Lempitsky V 2018 Deep Image Prior *Proc. IEEE Conf. Comput. Vis. Pattern Recognit.* pp 9446–54
13   Ulyanov D, Vedaldi A and Lempitsky V 2020 Deep Image Prior *Int. J. Comput. Vis.* **128** 1867–88
14   Gong K, Catana C, Qi J and Li Q 2019 PET image reconstruction using deep image prior *IEEE Trans. Med. Imaging* **38** 1655–65
15   Yokota T, Kawai K, Sakata M, Kimura Y and Hontani H 2019 Dynamic PET image reconstruction using nonnegative matrix factorization incorporated with deep image prior *Proc. IEEE/CVF Int. Conf. Comput. Vis.* pp 3126–35
16   Hashimoto F, Ote K and Onishi Y 2022 PET image reconstruction incorporating deep image prior and a forward projection model *IEEE Trans. Radiat. Plasma Med. Sci.* **6** 841–6
17   Sun H, Peng L, Zhang H, He Y, Cao S and Lu L 2021 Dynamic PET image denoising using deep image prior combined with regularization by denoising *IEEE Access* **9** 52378–92
18   Cui J, Gong K, Guo N, Wu C, Meng X, Kim K, Zheng K, Wu Z, Fu L, Xu B, Catana C, Qi J and Li Q 2019 PET image denoising using unsupervised deep learning *Eur. J. Nucl. Med. Mol. Imaging* **46** 2780–9
19   Hashimoto F, Ohba H, Ote K, Teramoto A and Tsukada H 2019 Dynamic PET image denoising using deep convolutional neural networks without prior training datasets *IEEE Access* **7** 96594–603
20   Onishi Y, Hashimoto F, Ote K, Ohba H, Ota R, Yoshikawa E and Ouchi Y 2021 Anatomical-guided attention enhances unsupervised PET image denoising performance *Med. Image Anal.* **74** 102226
21   Hashimoto F, Ohba H, Ote K, Kakimoto A, Tsukada H and Ouchi Y 2021 4D deep image prior: dynamic PET image denoising using an unsupervised four-dimensional branch convolutional neural network *Phys. Med. Biol.* **66** 015006 22 Cui J, Gong K, Guo N, Wu C, Kim K, Liu H and Li Q 2021 Populational and individual information based PET image denoising using conditional unsupervised learning *Phys. Med. Biol.* **66** 155001

23 Onishi Y, Hashimoto F, Ote K, Matsubara K and Ibaraki M 2024 Self-supervised pre-training for deep image prior-based robust PET image denoising *IEEE Trans. Radiat. Plasma Med. Sci.* **8** 348–56

24 Wang Y, Yu B, Wang L, Zu C, Lalush D S, Lin W and Zhou L 2018 3D conditional generative adversarial networks for high-quality PET image estimation at low dose *Neuroimage* **174** 550–62

25 Zhou L, Schaefferkoetter J D, Tham I W, Huang G and Yan J 2020 Supervised learning with CycleGAN for low-dose FDG PET image denoising *Med. Image Anal.* **65** 101770

26 Gong K, Johnson K, El Fakhri G, Li Q and Pan T 2024 PET image denoising based on denoising diffusion probabilistic model *Eur. J. Nucl. Med. Mol. Imaging* **51** 358–68

27 Zhu C, Byrd R H, Lu P and Nocedal J 1997 Algorithm 778: L-BFGS-B: fortran subroutines for large-scale bound-constrained optimization *ACM Trans. Math. Softw.* **23** 550–60

28 Akamatsu G, Takahashi M, Tashima H, Iwao Y, Yoshida E, Wakizaka H, Kumagai M, Yamashita T and Yamaya T 2022 Performance evaluation of VRAIN: a brain-dedicated PET with a hemispherical detector arrangement *Phys. Med. Biol.* **67** 225011

29 Takahashi M, Akamatsu G, Iwao Y, Tashima H, Yoshida E, Wakizaka H, Kumagai M, Yamashita T and Yamaya T 2022 Small nuclei identification with a hemispherical brain PET *EJNMMI Phys.* **9** 69

30 Pan S, Abouei E, Peng J, Qian J, Wynne J F, Wang T, Chang C-W, Roper J, Nye J A, Mao H and Yang X 2024 Full-dose whole-body PET synthesis from low-dose PET using high-efficiency denoising diffusion probabilistic model: PET consistency model *Med. Phys.* **51** 5468–78

31 Yu B, Ozdemir S, Dong Y, Shao W, Pan T, Shi K and Gong K 2025 Robust whole-body PET image denoising using 3D diffusion models: evaluation across various scanners, tracers, and dose levels Eur. J. Nucl. Med. Mol. Imaging doi: 10.1007/s00259-025-07122-4

32 Ote K, Hashimoto F, Onishi Y, Isobe T and Ouchi Y 2023 List-mode PET image reconstruction using deep image prior *IEEE Trans. Med. Imaging* **42** 1822–34